\documentclass{article}

\pdfoutput=1

\usepackage{jheppub}
\usepackage{float}
\usepackage{ulem}
\usepackage{mathrsfs}
\usepackage{natbib}
\usepackage{tikz}
\usepackage{setspace}
\usepackage{amsfonts}
\usepackage{amssymb}
\usepackage{amsmath}
\usepackage{esint}  
\usepackage{bbold}
\usepackage{epsfig}
\usepackage{physics}
\usepackage{latexsym}
\usepackage{color}
\usepackage{graphicx}
\usepackage{hyperref}
\usepackage{ulem}
\usepackage{bbm}

\usepackage[dvipsnames]{xcolor}
\DeclareMathAlphabet{\mathbbold}{U}{bbold}{m}{n}

\usepackage{parskip}
\usepackage{mathtools}

\definecolor{myDeepPurple}{RGB}{95, 5, 120}   
\definecolor{myDeepBlue}{RGB}{15, 35, 110}    
\definecolor{myDeepOrange}{RGB}{170, 60, 0}   
\hypersetup{
    colorlinks = true,
    citecolor  = myDeepPurple,
    linkcolor  = myDeepBlue,
    urlcolor   = myDeepOrange
}

\usepackage{nicefrac}
\usepackage{inputenc}
\usepackage{pstricks}
\usepackage{slashed}
\usepackage{multirow}
\usepackage{enumerate}
\usepackage{subcaption,comment}

\def\be{\begin{equation}}
\def\ee{\end{equation}}
\def\ba{\begin{eqnarray}}
\def\ea{\end{eqnarray}}

\def\qq{\qquad}

\def\IR{\relax{\rm I\kern-.18em R}}

\def\inv{^{\raise.0ex\hbox{${\scriptscriptstyle -}$}\kern-.05em 1}}

\allowdisplaybreaks

\definecolor{DCviolet}{RGB}{140, 43, 226}

\begin{document}

\title{Holographic Krylov spread complexity for a confining multi-charge AdS soliton }

\author[a]{Dimitrios Chatzis,}
\author[b]{Madison Hammond,  }
\author[b]{Ricardo T. Santamaria}
\author[b]{and Jonathan Whittle    }

\affiliation[a]{Instituto de F\'isica Te\'oretica, UNESP - Universidade Estadual Paulista R. Dr. Bento T. Ferraz 271, Bl. II, Sao Paulo 01140-070, SP, Brazil}

\affiliation[b]{Centre for Quantum Fields and Gravity, Department of Physics, Swansea University, Swansea SA2 8PP, United Kingdom}

\emailAdd{dchatzis@proton.me}
\emailAdd{m.hammond.2412736@swansea.ac.uk}
\emailAdd{ricardo.sta2718@gmail.com}
\emailAdd{jonathan.whittle@swansea.ac.uk}

\abstract{
We study holographic spread complexity for a Type IIB solution dual to a deformation of $\mathcal{N} = 4$ SYM that flows to a confining (2+1)-dimensional supersymmetric theory in the IR. We use the proposal that relates the rate of change of spread complexity of local operators with the proper momentum of an infalling probe particle in the bulk. We analyse two cases: a probe particle falling radially and a probe falling radially while rotating with conserved angular momentum. We elaborate on how to define the proper coordinate in the presence of a conserved Noether charge using the Routhian. The trajectories for the probes were studied numerically, revealing an oscillatory behaviour of the complexity.

}

\maketitle

\flushbottom

\section{Introduction}

Quantum complexity has emerged as a useful framework for characterising the growth of quantum states and operators under time evolution. 
A variety of notions of complexity have been developed, including circuit complexity, Nielsen complexity, spread complexity and Krylov complexity, each capturing different aspects of quantum evolution. For reviews of recent developments, see \cite{Nandy:2024evd,Rabinovici:2025otw,Baiguera:2025dkc}. 

The holographic correspondence provides a natural setting in which to investigate the gravitational interpretation of quantum complexity. In particular, the complexity=volume and complexity=action proposals \cite{Susskind:2014rva,Brown:2015bva} relate the growth of quantum complexity to geometric quantities associated with the dual bulk spacetime. These developments suggest that gravitational dynamics may encode aspects of quantum information growth in strongly coupled field theories, providing a new perspective on the relation between geometry and quantum dynamics.

In this work, we focus on spread complexity. Given a normalised initial state $|K_0 \rangle$, the Krylov basis is generated through repeated action of the Hamiltonian using the Lanczos algorithm \cite{Lanczos:1950zz}, which constructs an orthonormal basis that \textit{minimises} the spread of such a state under time evolution. The Hamiltonian acts tri-diagonally,
\begin{equation}
H|K_n\rangle=b_{n+1}|K_{n+1}\rangle+a_n|K_n\rangle+b_n|K_{n-1}\rangle.
    \label{tridiag}
\end{equation}
This basis is the Krylov basis, $\ket{K_{n}}$, which defines an effective one-dimensional Krylov chain, where the Lanczos coefficients, $\{a_n, b_n \}$, determine the hopping amplitudes between neighbouring sites. The growth of the operator along this chain provides a measure of complexity, with the Lanczos coefficients encoding information about the dynamics. 

The Lanczos coefficients can be obtained from the moments of the survival amplitude $S(t) = \langle{K_{0}}|e^{-iHt}|K_{0}\rangle$ \cite{Balasubramanian:2022tpr,Caputa:2023vyr,Balasubramanian:2025xkj,Caputa:2025mii}. At the level of the wavefunction $\phi_{n}(t) = \langle{K_{n}}|e^{-iHt}|K_{0}\rangle$, equation \eqref{tridiag} is equivalent to a discrete Schrödinger--like equation
\begin{equation}
 i\,\partial_t\phi_n(t)=a_n\phi_n(t)+b_{n+1}\phi_{n+1}(t)
 +b_n\phi_{n-1}(t),
 \qquad \phi_n(0)=\delta_{n0} \, .
 \label{hopping}
\end{equation}
This equation defines a distribution that quantifies the probability of evolving to a given state in the chain. In this way spread complexity is defined as the average position along the Krylov chain
\begin{equation}
    \mathcal{C}(t) = \sum_{n} n \,  |\phi_{n}(t)|^{2} \, .
\end{equation}
The early time behaviour is determined from the equation \eqref{hopping}, leading to \cite{Caputa:2025mii,Caputa:2025ozd,Huh:2023jxt}
\begin{equation}
    {\cal C}(t) = b_1^2t^2+b_1^2\left[
 \frac{b_2^2}{6}-\frac{b_1^2}{3}
 -\frac{(a_1-a_0)^2}{12}\right]t^4+{\cal O}(t^6),
 \label{earlytime}
\end{equation}
such that $\mathcal{C}(0) = 0, \; \mathcal{\dot{C}}(0)=0$. Furthermore, $\mathcal{C}(t)$ is an even function \cite{Muck:2026top, Aguilar-Gutierrez:2025kmw}. This early time behaviour is also true in the presence of conserved charges, for each of the fixed charge sectors in the Hilbert space \cite{Caputa:2025ozd}.

A holographic description of Krylov spread complexity was proposed in \cite{Caputa:2024sux}, where the evolution of the Krylov state is mapped to the motion of a massive probe particle in an asymptotically AdS spacetime. According to this conjecture, the growth rate of Krylov complexity is related to the proper momentum of the particle,
\begin{equation}
\mathcal{\dot{C}}(t) \simeq  \Lambda \, P_y(t) \, ,
\end{equation}
for $\Lambda$ a positive constant and $y$ the proper coordinate, providing a geometric description of operator growth in the dual strongly coupled field theory. This proposal offers a framework in which the behaviour of Krylov complexity can be studied directly from gravitational backgrounds.

Recently, spread complexity has been studied holographically in a range of setups, including confining theories, conformal theories and matrix models \cite{Nunez:2026vhw,Nunez:2026kwr,Fatemiabhari:2026rob,Fatemiabhari:2026goj,Fatemiabhari:2025poq,Fatemiabhari:2025usn,Zoakos:2026obl,Fatemiabhari:2025cyy,Fatemiabhari:2026eem}.
Moreover, spread complexity of extended objects has also been studied recently \cite{Chatzis:2026ekd,Nastase:2026lhz,Chatzis:2026oou}, generalising the proposal of the massive particle.

The papers \cite{Chatzis:2026ekd,Nastase:2026lhz,Nunez:2026vhw} show how to define the proper coordinate $y$ when the probe has a conserved Noether charge. They implement the Routhian, which is a Legendre transformation of the Lagrangian to a fixed charge sector, mimicking the fixed charge sectors on the Hilbert space of the dual field theory. This prescription allows us to obtain the correct early time behaviour in \eqref{earlytime}.

The gravity background we consider here belongs to a wider class of confining backgrounds which have been studied in detail in the recent literature \cite{Anabalon:2021tua,Anabalon:2024che,Anabalon:2026yxk,Anabalon:2022aig,Anabalon:2024qhf,Chatzis:2024kdu,Chatzis:2024top,Chatzis:2025hek,Macpherson:2025pqi,Macpherson:2024qfi,Chatzis:2025dnu,Chatzis:2025wfv,Hammond:2026lle,Nunez:2023nnl,Nunez:2023xgl,Kumar:2024pcz}. The purpose of this work is to investigate holographic spread complexity in a class of smooth AdS-soliton solutions arising from gauged supergravity. These backgrounds describe the vacuum structure of $\mathcal{N}=4$ super Yang--Mills theory compactified on a spatial circle of perimeter $L_\varphi$. We focus on the holographic construction of \cite{Anabalon:2026yxk} which provides a complete description of this system, including both supersymmetric and non-supersymmetric AdS-soliton solutions which are asymptotically AdS$_5$ and smooth in the infrared. The dual field theory contains vacuum expectation values for scalar bilinears together with three independent current sources associated with the Cartan subgroup $U(1)^3\subset SO(6)_R$, which in the SUSY case reduces to one $U(1)$ R-symmetry. These sources admit a geometric interpretation as twists along the isometry directions of the internal $S^5$.

In this paper we add to the growing body of literature on complexity in confining systems, applying the holographic complexity proposal of \cite{Caputa:2024sux} to the AdS-soliton geometry of \cite{Anabalon:2026yxk}. This complements previous studies \cite{Anabalon:2024che} which focus on a subregion of the full moduli space. By analysing the geodesic motion of the dual massive particle, we study the dependence of spread complexity on the parameters controlling the vacuum structure and explore how features of the bulk geometry are reflected in the growth of complexity. By adopting the Routhian prescription we contribute to the mounting evidence that this is the correct formalism in which to compute complexity in systems possessing conserved charges, and show that doing so reproduces the expected early-time behaviour for the complexity.

The rest of the paper is organised as follows. Section \ref{sec: background} reviews the Type IIB background. Section \ref{sec: motivation routhian} provides a brief motivation on how to define the proper coordinate in the presence of conserved charges. In Section \ref{sec: embedding 1} we study the geodesic of a falling particle with no angular momentum and calculate its complexity. Section \ref{sec: emb II} extends this to the example of a probe particle falling radially with angular momentum in the confining circle direction and one of the internal angles. In section~\ref{sec:field_theory} we identify the initial state $|K_0\rangle$ in the latter case and compute holographically the first Lanczos coefficient $b_1$ as well as the second moment of the return amplitude. Section \ref{conclusions} gives a summary and conclusions. 

\section{The background}\label{sec: background}

    We consider the background presented in \cite{Anabalon:2026yxk}, obtained via an uplift to 10-dimensional Type IIB supergravity from a solution to the gauged STU model in 5 dimensions \cite{Behrndt:1998jd,Cvetic:1999xp}.
    
    The bosonic field content of the gauged STU model contains a metric, two scalars $(\Phi_1, \Phi_2)$, and three $U(1)$ gauge fields $A^i$ for which the two-form field strengths are given by $F^i = \mathrm{d} A^i$. The 5-dimensional action is
    \begin{equation}
        S_{0} = \frac{1}{2\kappa} \int \mathrm{d}^{5} x\sqrt{-g} \left( R - \frac{(\partial \Phi_{1})^{2}}{2} - \frac{(\partial \Phi_{2})^{2}}{2}
        + \sum_{i=1}^{3} 4 L^{-2} X_{i}^{-1} - \frac{1}{4} X_{i}^{-2} (F^{i})^{2}
        + \frac{1}{4} \epsilon^{\mu \nu \rho \sigma \lambda}  A_{\mu}^{1} F_{\nu \rho}^{2} F_{\sigma \lambda}^{3} \right) \, ,
    \end{equation}
    where
    \begin{equation}
        X_i = \mathrm{e}^{-\frac{1}{2} \vec{a}_i \cdot \vec{\Phi}} \qq \mathrm{with} \qq \vec{a}_1 = \left( \frac{2}{\sqrt{6}}, \sqrt{2} \right), \quad
        \vec{a}_2 = \left( \frac{2}{\sqrt{6}}, -\sqrt{2} \right), \quad
        \vec{a}_3 = \left( -\frac{4}{\sqrt{6}}, 0 \right).
    \end{equation}
   We use the coordinates $(t,x_1,x_2,\varphi,r,\theta,\psi,\phi_1,\phi_2,\phi_3)$, in which the metric is given by
    \begin{equation}\label{eq:metric_10d}
        \mathrm{d} s_{10}^2= \tilde{\Delta}^{1 / 2} \frac{r^2}{L^2} H(r)^{1 / 3}\left( \mathrm{d} x_{1,2}^2 +\frac{L^2}{r^2} \frac{f(r)}{H(r)} \mathrm{d} \varphi^2+\frac{L^2}{r^2} \frac{\mathrm{~d} r^2}{f(r)} \right)  +L^2 \tilde{\Delta}^{-1 / 2} \sum_{i=1}^3 X_i^{-1}\left[\mathrm{~d} \mu_i^2+\mu_i^2\left(\mathrm{~d} \phi_i+\frac{1}{L} A_i\right)^2\right]
    \end{equation}
    which is a deformed AdS\textsubscript{5}$\times S^5$ geometry, where we have chosen the following $\mu_i$ to parametrise the $S^5$
    \begin{align*}
        \mu_1=\cos \theta \cos \psi, \qq \mu_2=\cos \theta \sin \psi, \qq \mu_3=\sin \theta.
    \end{align*}
    The warp factors are given by
    \begin{align}
        & f(r) = \frac{r^2}{L^2} H(r) - \frac{Q^{2}}{r^4} , \qq H(r) = H_1(r)H_2(r)H_3(r) \qq \mathrm{where} \qq H_i(r) = 1 + \frac{q_i^2}{r^2}, \qq  \\
        &\tilde{\Delta}=\sum_i X_i \mu_i^2 =\cos^{2}\theta\left( X_{1}\cos^{2}\psi + X_{2}\sin^{2}\psi \right) + X_{3}\sin^{2}\theta.
    \end{align}
    The background is asymptotically $\mathrm{AdS}_5 \times S^5$ as $r\rightarrow\infty$. Away from the boundary, the warp factors have a non-trivial radial dependence and the compact $\varphi$ circle shrinks to zero at $r=r_*$, where the geometry ends smoothly. $r_*$ is defined by $f(r_*) = 0$. Requiring that the space caps off smoothly at $r=r_*$ results in the following regularity condition on the period $L_\varphi$ of $\varphi$:
    \begin{equation}\label{eqn: regularity condition}
        L_\varphi = \frac{4\pi H(r_*)^{1/2}}{| f'(r_*) |},
    \end{equation}
    where a prime represents differentiation with respect to $r$, and we can use $L_\varphi$ as a measure of the compactification scale (the KK modes will have a mass $ \sim L_\varphi^{-1}$). The three gauge fields are
    \begin{equation}
        A_i = f_i(r) \ \dd \varphi \qq \mathrm{where} \qq f_i(r) = Q\left(\frac{1}{r^2 H_i(r)} - \frac{1}{r_*^2 H_i(r_*)}\right),
    \end{equation}
    the scalars $(\Phi_1, \Phi_2)$ give the following form to the $X_i$ functions
    \begin{equation}
        X_1^3=\frac{H_2(r) H_3(r)}{H_1^2(r)}, \qq X_2^3=\frac{H_1(r) H_3(r)}{H_2^2(r)}, \qq X_3^3=\frac{H_1(r) H_2(r)}{H_3^2(r)},
    \end{equation}
    and finally, the five-form field strength is given by
    \begin{align}\label{eq:flux}
        F_5=  G_5+\star G_5 \qq \mathrm{where} \qq G_5= \ & \frac{2}{L} \sum_i\left(X_i^2 \mu_i^2-\tilde{\Delta} X_i\right) \star_5 \mathbb{1} + \frac{L}{2} \sum_i X_i^{-1} \star_5 \mathrm{~d} X_i \wedge \mathrm{~d}\mu_i^2 \nonumber \\ & +  L^2 \sum_i X_i^{-2} \mu_i \mathrm{~d} \mu_i \wedge\left(\mathrm{~d} \phi_i+\frac{1}{L} A_i\right) \wedge \star_5 F_i
    \end{align}
    where the Hodge dual in 10-dimensions is denoted by $\star$, the Hodge dual in five dimensions with respect to the 5-dimensional metric is $\star_5$, and we adopt the conventions of \cite{Anabalon:2026yxk} for the Hodge dual of a $p$-form $F_p$.
    The supersymmetric type IIB uplifted solution of \cite{Anabalon:2026yxk} depends on the parameters $(q_1,q_2,q_3,Q)$\footnote{The parameter $M$ in \cite{Anabalon:2026yxk} is set to zero.}, and contains a metric and a five-form field strength, as well as the scalars and gauge fields from the 5-dimensional solution.
    The dual theory is a deformation of $\mathcal{N} = 4$ SYM due to the the circle compactification $\mathbb{R}^{1,2} \times S^{1}$ and the twist performed by the gauge fields $A_i$, yielding non-zero R-symmetry holonomies around the $S^1$. In the IR the theory flows to a gapped $(2+1)$-dimensional QFT, which in the SUSY case, preserves $4$ supercharges. In this work, we study the spread complexity for local operators of this theory that are dual to probe particle excitations in the bulk -- the evolution of the operator is mapped to geodesic motion in the gravity side. These operators can be charged under the R-symmetry, which in the bulk corresponds to probe particles falling with a profile in one of the internal coordinates. \\ \\ In the following sections we present two geodesics for a particle in the background \eqref{eq:metric_10d}-\eqref{eq:flux}. The first describes a radially infalling particle, while the second embedding includes rotations along the confining circle direction $\varphi$ and one of the three $\text{U}(1)$ isometries of the $S^5$. For the following calculations we choose the change of coordinates $\frac{r}{L}=\frac{L}{\zeta}$, so that the boundary of the geometry now lies at $\zeta = 0$, and the end of space is at $\zeta_* = \frac{L^2}{r_*}$. We reference explicitly the metric in the following form for the sake of convenience 
    \begin{equation}\label{eqn: metric in z}
        \begin{split}
            \dd s^2_{10} & =\frac{L^2H^{1/3}\tilde{\Delta}^{1/2}}{\zeta^2}\left( -\dd t^2 + \dd x_1^2 + \dd x_2^2 + \frac{L^2}{\zeta^2} \frac{\dd \zeta^2}{f} + \frac{\zeta^2}{L^2} \frac{f \dd \varphi^2 }{H}\right) + \\
            &\frac{L^2}{\tilde{\Delta}^{1/2}}\left[ \frac{\dd \mu_1^2+\mu_1^2 (\dd \phi_1+L^{-1}f_1 \dd \varphi)^2}{X_1}+\frac{\dd \mu_2^2+\mu_2^2 (\dd \phi_2+L^{-1}f_2 \dd \varphi)^2}{X_2} +\frac{\dd \mu_3^2+\mu_3^2 (\dd \phi_3+L^{-1}f_3 \dd \varphi)^2}{X_3}  \right]. 
        \end{split}
    \end{equation}
As mentioned in \cite{Anabalon:2026yxk}, moving in the parameter space characterising the solution \eqref{eq:metric_10d}-\eqref{eq:flux} one can recover the solutions of \cite{Anabalon:2021tua} and \cite{Anabalon:2024che}. The single charge solution of Anabal\'on and Ross \cite{Anabalon:2021tua} can be reached by setting $q_i=0$. In this case we have $\tilde{\Delta}=X_i=H_i=1$ and the SUSY solution is characterised by $Q$. Less obviously, we can make contact with the two parameter solution of \cite{Anabalon:2024che} which includes a single excited scalar, for which we set $q_1=q_2\equiv Q_1$, $q_3=Q_2$ leading to $X_1=X_2=X_3\equiv X=H_3/H_1$. Here we will focus on the SUSY solution.
In the next section, we motivate the Routhian procedure for calculating holographic Krylov complexity for when the system has a conserved Noether charge. 

\section{Routhian description} \label{sec: motivation routhian}
    Previously, the proper momentum used to calculate the holographic Krylov complexity had been defined from the Lagrangian. In \cite{Nastase:2026lhz,Chatzis:2026ekd}, it was suggested that for dynamics in backgrounds possessing continuous isometries the appropriate quantity is instead the Routhian, obtained by performing a Legendre transformation with respect to the coordinates with conserved charges. In this section we explain the differences for defining the proper coordinate between the Lagrangian and the Routhian prescriptions.

    In previous works \cite{Fatemiabhari:2026rob,Fatemiabhari:2026goj,Fatemiabhari:2025cyy,Fatemiabhari:2025poq,Fatemiabhari:2025usn}, the proper coordinate $\tilde{y}$ used to define the proper momentum such that $\dot{\mathcal{C}} \sim P_{\tilde{y}}$, was defined in the Lagrangian as
    \begin{equation}\label{eqn: L prop form}
    	\mathcal{L} = -m \sqrt{\tilde{h}(\zeta)^2 - \dot{\tilde{y}}^2}
    \end{equation}
    where $\tilde{h}(\zeta)$ is a function of the radial coordinate, but importantly containing no time derivatives, all of which are absorbed into the definition of $\dot{\tilde{y}}$. The proper momentum is then $P_{\tilde{y}} = \frac{\partial \mathcal{L} }{ \partial \dot{\tilde{y}}}$. This definition however does not capture the expected behaviour of the complexity in the presence of a conserved charge \cite{Nastase:2026lhz,Chatzis:2026ekd,Chatzis:2026oou}. When the system has a conserved charge, the early-time behaviour of the proper momentum derived from the Lagrangian is typically $P_{\tilde{y}} \sim k + \mathcal{O}(t^2)$, where $k$ is a constant proportional to the conserved charge, which results in a complexity with early-time behaviour of $\mathcal{C} \sim t + \mathcal{O}(t^3)$. Krylov complexity is known to admit an even-powered expansion in $t$ (see eg. \cite{Muck:2026top, Aguilar-Gutierrez:2025kmw}) so it seems necessary to search for a different method for defining the proper momentum.
    Recently it has been proposed \cite{Nastase:2026lhz,Chatzis:2026ekd} that the correct treatment when the system possesses a conserved charge is to work in a fixed-charge sector, which means instead defining the proper momentum from the Routhian, an analogous object to the Lagrangian but differing by a Legendre transform. For example in Section \ref{sec: emb II} we will encounter a system with conserved angular momentum $J = \frac{\partial \mathcal{L} }{\partial \dot{\varphi}}$ along a cyclic direction $\varphi$. The Routhian for this system is obtained by:
    \begin{equation}
    	\mathcal{R}(\zeta, \dot{\zeta}) = J \dot{\varphi}(\zeta, \dot{\zeta}) - \mathcal{L}(\zeta, \dot{\zeta}).
    \end{equation}
    Then the proper coordinate is defined such that the Routhian can be written in an analogous form to that of the Lagrangian in equation \eqref{eqn: L prop form},
    \begin{align}
        \mathcal{R} = \sqrt{h(\zeta)^2 - \dot{y}^2}
    \end{align}
    where once again $h(\zeta)$ is a function of the radial coordinate, but importantly containing no time derivatives -- again all time derivatives of coordinates are absorbed into the definition of $\dot{y}$.
    The proper momentum is then given by $P_y = \frac{\partial \mathcal{R} }{ \partial \dot{y}}$. In all examples considered, the Routhian prescription results in proper momenta with early-time behaviour like $P_y \sim t + \mathcal{O}(t^3)$ which means that the complexity goes like $\mathcal{C} \sim t^2 + \mathcal{O}(t^4)$, thus reproducing the expected behaviour. In the limit $J \rightarrow 0$ the Routhian prescription reproduces the results derived from the Lagrangian.
    
    We first consider an embedding with no conserved charges, and as a result the Lagrangian prescription is appropriate. Following this we consider an embedding which possesses a conserved charge of the type described above, and will therefore adopt the Routhian prescription.
\section{Radially falling particle without angular momentum}\label{sec: embedding 1}
    We now consider a massive probe particle falling in the background described by \eqref{eqn: metric in z}. The simplest embedding consistent with the equations of motion is the probe falling radially, with profile $\zeta = \zeta(t)$ and all other coordinates constant. For consistency of the equations of motion we require also that $\theta = \psi = 0$. The induced metric on the particle world-line is then given by
    \begin{equation}
        \dd s^2_{\text{ind}}=\frac{H^{1/3}(\zeta)X_1(\zeta)^{1/2}}{f(\zeta) \zeta^4} \left( -L^2 f(\zeta) \zeta^2+L^4 \dot{\zeta}^2 \right)\dd t^2 \, .
    \end{equation}
    We define the function 
    \begin{equation}\label{G}
    G(\zeta) = H^{1/6}(\zeta) X_1^{1/4}(\zeta).
    \end{equation}
    This allows us to write the action for the particle in the following form:
    \begin{equation}
        \begin{split}
            S = \int \dd t \ {\cal L}_{\rm I} \qq \mathrm{where} \qq {\cal L}_{\rm{I}}= -m \, G(\zeta) \sqrt{\frac{L^{2}}{\zeta^{2}}-\frac{L^{4}}{\zeta^{4}}\frac{ \dot{\zeta}^2}{f(\zeta)}}.
        \end{split}
    \end{equation}
    From the conserved Hamiltonian $\mathcal{H}_{\rm I} = P_{\zeta}\dot{\zeta} - \mathcal{L}_{\rm I}$, we obtain the equation  
    \begin{equation}\label{eqn: zetadot I}
        \dot{\zeta} = \pm \sqrt{f(\zeta)\left(\frac{\zeta^{2}}{L^{2}}-\frac{m^{2}}{\mathcal{H}_{\rm I}^{2}}G(\zeta)^{2}  \right)} \, .
    \end{equation}

    Solving this differential equation for $\zeta(t)$ gives us the trajectory of the probe particle. We start the particle at a large radial position $r_{\text{UV}}$ which corresponds to a small $\zeta$ given by $\zeta_{\text{UV}} = \frac{L^2}{r_{\text{UV}}}$. The boundary conditions for which we solve the equation are $\zeta(t = 0) = \zeta_{\text{UV}}$ and $\dot{\zeta} (t = 0) = 0$. By substitution into \eqref{eqn: zetadot I} we can express $\mathcal{H}_I$ in terms of $\zeta_{UV}$, yielding
    \begin{equation}\label{eqn: zetadotUV I}
        \dot{\zeta} = \pm \frac{\zeta}{L} \sqrt{f(\zeta)\left[1-\left(\frac{G(\zeta)\zeta_{\text{UV}}}{G(\zeta_{\text{UV}})\zeta}\right)^{2}  \right]} \, .
    \end{equation}
    Notice that it is now explicit that $\dot{\zeta}(t = 0) = 0$. In addition, we always have that $\zeta_{\text{UV}} \leq \zeta(t)$, and as a consequence the argument of the square root is well defined. At the end of space $f(\zeta_*) = 0$ and as a result the velocity in $\zeta$ vanishes - this implies that the particle stops at the end of the space and bounces back towards $\zeta_{\text{UV}}$, thus explaining the two possible signs in equation \eqref{eqn: zetadotUV I}. 
    
\begin{figure}[t!]
    	\centering
    	\includegraphics[width = 3.5in]{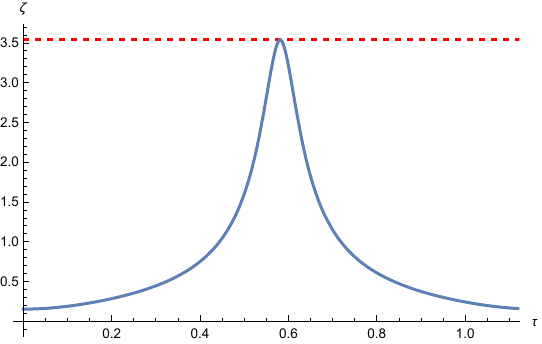}
    	\caption{Trajectory for the probe. The set of parameters used were $(q_1,q_2,q_3,Q^2, \mathcal{H}_{\rm I}) = (1,2,3,40,7)$ and we set $L=1$. The dashed line represents the end of the space $\zeta_{*}$. We plot one period of oscillation, but the pattern continues with the probe particle falling towards the end of space, and then bouncing back and returning to its starting position at $\zeta_{\text{UV}}$.}
        \label{zt1}
    \end{figure}

    Analytic solutions to \eqref{eqn: zetadotUV I} are highly non-linear, so we solve the equation numerically over the range $\zeta(t) \in [\zeta_{\text{UV}}, \zeta_{*}]$ after specifying the parameters $(q_1, q_2, q_3, Q^2, \mathcal{H}_I)$. Solving $\dot{\zeta} = 0$ in equation \eqref{eqn: zetadot I} gives us two stationary point solutions, one being $\zeta_{\text{UV}}$ and the other being the turning point, at $\zeta_*$. After choosing values for the parameters, this gives us the range over which to solve the differential equation (and attempting to look for a solution outside of this range returns complex-valued embeddings). Then after solving \eqref{eqn: zetadot I} numerically, subject to the aforementioned boundary conditions, we find that the particle falls, reaches the end of space, and then bounces back towards $\zeta_{\text{UV}}$. See Figure \ref{zt1} for a typical embedding.

    The requirement that $f(\zeta_*) = 0$ has real solutions restricts us to consider only embeddings with $L^2 Q^2 \geq q_1^2 q_2^2 q_3^2 $.

    At early times we can expand equation \eqref{eqn: zetadotUV I} and solve order-by-order in $t$, yielding
    \begin{equation}\label{eqn: early time zeta I}
        \zeta(t) \approx \zeta_{\text{UV}} + \frac{ f(\zeta_{\text{UV}}) \left(G(\zeta_{\text{UV}})-G'(\zeta_{\text{UV}}) \zeta_{\text{UV}} \right)}{2 G(\zeta_{\text{UV}})} \frac{\zeta_{\text{UV}} t^{2}}{L^{2}} + \mathcal{O}(t^{4}) \qq \text{for} \qq t \to 0
    \end{equation}
    where primes denote differentiation with respect to $\zeta$. Note in particular that there is no term linear in $t$. This shows close agreement with the numerically-obtained trajectory of Figure \ref{zt1} at early times.

    Similarly we can expand equation \eqref{eqn: zetadotUV I} around the turning point (which for this embedding is the end of space $\zeta_*$), yielding
    \begin{equation}
        \zeta(t) \approx \zeta_* + \frac{(\mathcal{H}^2 \zeta_*^2 - G(\zeta_*^2)) f'(\zeta_*)}{4 \mathcal{H}^2} (t - t_*)^2 + \mathcal{O}(t-t_{*})^{4} \qq \mathrm{for} \qq t \approx t_*
    \end{equation}
    where we define $t_*$ such that $\zeta(t_*) = \zeta_*$. Again we observe only terms even-powered in $t$.
    
    In the next subsection we compute the holographic complexity for this probe and explore the asymptotic behaviour using the above expansion.

    \subsection{Complexity}\label{complexityzeta}
    
\begin{figure}[t!]
    	\centering
    	\includegraphics[width = 6in]{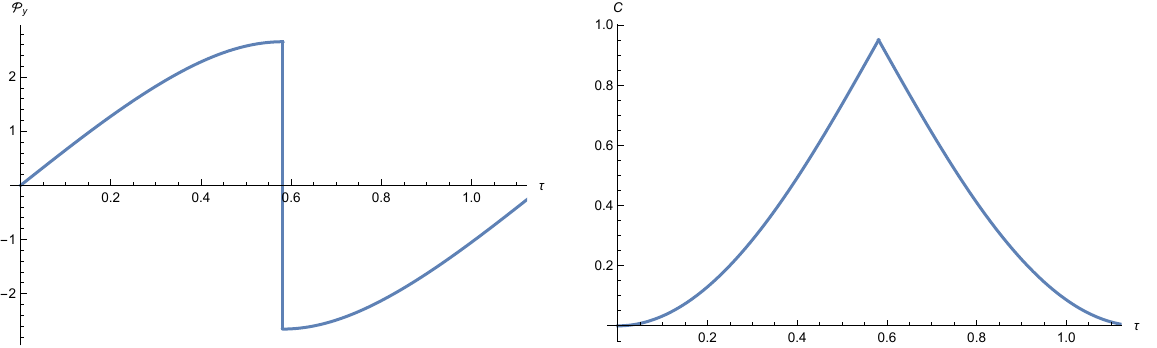}
    	\caption{Proper momentum and complexity associated with the particle trajectory in Figure \ref{zt1}, for the same choice of parameters.}
        \label{complexityzt}
    \end{figure}

    Since there are no conserved charges, the proper momentum can be defined from the Lagrangian or the Routhian (since they are identical in this case). The definition of the proper coordinate is
    \begin{equation}
        \dot{y} = \text{sign}(\dot{\zeta}) \,  G(\zeta) \frac{L^{2}|\dot{\zeta}|}{\zeta^{2} f(\zeta)^{1/2}},
    \end{equation}
   which allows us to rewrite the Lagrangian in the following way:
    \begin{equation}
        \mathcal{L}_{\rm I} = -m \sqrt{G(\zeta)^{2}\frac{L^{2}}{\zeta^{2}}-\dot{y}^{2} }.
    \end{equation}
   The sign of the proper coordinate $\dot{y}$ depends on the sign of $\dot{\zeta}$, meaning whether the particle is falling or bouncing back. With this definition, the proper momentum $P_{y} = \frac{\partial \mathcal{L}_{\rm I}}{\partial \dot{y}}$ is given by
   \begin{equation}
       P_{y} = \text{sign}(\dot{\zeta}) \, \frac{m |\dot{y}|}{\sqrt{G(\zeta)^{2}\frac{L^{2}}{\zeta^{2}}-\dot{y}^{2} } } = \text{sign}(\dot{\zeta}) \, \frac{m |\dot{\zeta}|}{\sqrt{ \frac{f(\zeta)\zeta^{2}}{L^{2}}-\dot{\zeta}^{2}}} \, .
       \label{proper0}
   \end{equation}
    Notice that for this case $[P_{y}] = \frac{1}{\text{length}}$. Additionally, if we employ equations \eqref{eqn: zetadot I} and \eqref{eqn: zetadotUV I} we can obtain an explicit expression for the proper momentum in terms of $\zeta$:
    \begin{equation}\label{eqn: proper momemtum I G}
        P_{y} =   \text{sign}(\dot{\zeta}) \,  m \,  \sqrt{\left( \frac{\mathcal{H}\zeta}{G(\zeta) \, m L }\right)^{2}-1} =  \text{sign}(\dot{\zeta}) \,  m \, \sqrt{\left(\frac{G(\zeta_{\text{UV}})\zeta}{G(\zeta)\zeta_{\text{UV}}}\right)^{2}-1} \, .
    \end{equation}
    Into this last expression we substitute our (numerically obtained) trajectories for $\zeta(t)$. Then the complexity $\dot{\mathcal{C}} \sim P_{y} $ is obtained by integration, which again must be done numerically after substituting in the trajectories. See Figure \ref{complexityzt} for plots of the proper momentum and complexity associated to this embedding for a generic choice of the parameters. The proper momentum has a discontinuity when the particle bounces off the end of space, and as a result the complexity is kinked at this point. We have plotted only one period of oscillation but as with the particle trajectory this pattern repeats as we look to larger times. Oscillating behaviour of the complexity is characteristic (but not diagnostic \cite{Nunez:2026vhw}) for confining systems.

    Another aspect worth exploring is the asymptotic (early time) behaviour of the proper momentum -- using the series expansion for $\zeta(t)$ in \eqref{eqn: early time zeta I} and substituting into \eqref{eqn: proper momemtum I G} we obtain
    \begin{equation}
        P_{y}(t) \approx m \frac{f(\zeta_{\text{UV}})^{1/2} \left(G(\zeta_{\text{UV}})-G'(\zeta_{\text{UV}}) \zeta_{\text{UV}} \right)}{L \, G(\zeta_{\text{UV}})} t + \mathcal{O}(t^{3}) \quad \text{for} \quad t \to 0 \, .
    \end{equation}
    Importantly we find that the proper momentum is linear to leading order in $t$, which means that the associated complexity will be quadratic to leading order. 
    
    The fact that $P_y$ has a step when $\zeta = \zeta_*$ prevents us from obtaining an expansion at this point. See Figure \ref{complexityzt}.
    
    In the next section, we will consider an embedding of the particle which now extends to the internal space.
   
\section{Radially falling particle with angular momentum}\label{sec: emb II}

    We now consider a more interesting embedding, corresponding to a particle rotating in $\varphi$. Consistency of the equations of motion requires the particle to also rotate in one of the $S^5$ directions $\phi_i$. Let us propose the following ansatz
    \begin{equation}
       \zeta=\zeta (t), \quad \varphi=\varphi (t),\quad \phi_1=\phi_1(t),  
    \end{equation}
    where compatibility with the equations of motion requires us to set
    \begin{equation}
         \dot{\phi}_1(t) = - \frac{1}{L} f_1(\zeta) \dot{\varphi}(t) \quad \text{and}\quad \theta=0=\psi \, .
    \end{equation}
     For this configuration, we have the following induced metric
    \begin{equation}
        \dd s^2_{\text{ind}}=\frac{X_1^{1/2}}{f H^{2/3} \zeta^4}\left[ L^2 H \left( -f \zeta^2 + L^2 \dot{\zeta}^2 \right)+ f^2 \zeta^4 \dot{\varphi}^2 \right]\dd t^2 \, .
    \end{equation}
    Then the Lagrangian is given by
    \begin{equation}\label{eqn: Lagrangian_II}
        {\cal L}_{\rm II} = -m\,  G(\zeta) \sqrt{\frac{L^{2}}{\zeta^{2}}-\frac{L^{4}}{\zeta^{4}}\frac{ \dot{\zeta}^2}{f(\zeta)} - \frac{f(\zeta)}{H(\zeta)} \dot{\varphi}^{2}} \, .
    \end{equation}
    Unlike the non-rotating case, there now exists a conserved quantity associated with $\varphi$, namely its canonical momentum:
    \begin{equation}
        J \equiv \frac{\partial \mathcal{L}_{\rm II}}{\partial \dot{\varphi}} = -\frac{m^2 f(\zeta) G(\zeta)^2 }{H(\zeta) \mathcal{L}_{\rm II}} \dot{\varphi} \, .
    \end{equation}
    This means that we need to use the Routhian prescription. Using the above expression for $J$, we can rearrange for $\dot{\varphi} = \dot{\varphi}(\zeta, \dot{\zeta}; J)$ which we then use to compute the Routhian, given by the Legendre transform
    \begin{align}\label{routhianj}
        \mathcal{R} = J \dot{\varphi}(\zeta, \dot{\zeta}) - \mathcal{L}_{\rm II}(\zeta, \dot{\zeta}; J) &= \sqrt{\frac{m^{2}f(\zeta)G(\zeta)^{2} + J^{2}H(\zeta)}{f(\zeta)}\left( \frac{L^{2}}{\zeta^{2}}-\frac{L^{4}}{\zeta^{4}}\frac{ \dot{\zeta}^2}{f(\zeta)}\right)} \\
        &\equiv \sqrt{g(\zeta)^{2}\left( \frac{L^{2}}{\zeta^{2}}-\frac{L^{4}}{\zeta^{4}}\frac{ \dot{\zeta}^2}{f(\zeta)}\right)} \qq \mathrm{where} \qq g(\zeta)^2 =  \frac{m^{2}f(\zeta)G(\zeta)^{2} + J^{2}H(\zeta)}{f(\zeta)}\nonumber.
    \end{align}
    From the Routhian, we can obtain a conserved Hamiltonian defined by $\mathcal{H}_{\rm II} = \frac{\partial {\cal R}}{\partial\dot{\zeta}}\dot{\zeta} - \mathcal{R} $. Note that in general this will be different to the Hamiltonian defined via the Lagrangian. Rearranging for $\dot{\zeta}$ in terms of $\mathcal{H}_{\rm II}$ we find:
  
    \begin{equation}\label{eqn: zetadot II}
        \dot{\zeta} = \pm \sqrt{f(\zeta)\left(\frac{\zeta^{2}}{L^{2}}-\frac{m^{2}}{\mathcal{H}_{\rm II}^{2}}G(\zeta)^{2} \right)-\frac{J^{2}}{\mathcal{H}_{\rm II}^{2}}H(\zeta) } \,\, .
    \end{equation}
    Notice that in the limit $J\to 0$ we recover equation \eqref{eqn: zetadot I}. Moreover, note that in this case the radial point at which the velocity vanishes is no longer the root of the function $f(\zeta)$, namely $\zeta=\zeta_*$, but rather a new turning point $\zeta_0<\zeta_*$.
    
\begin{figure}[t!]
    	\centering
    	\includegraphics[width = 4in]{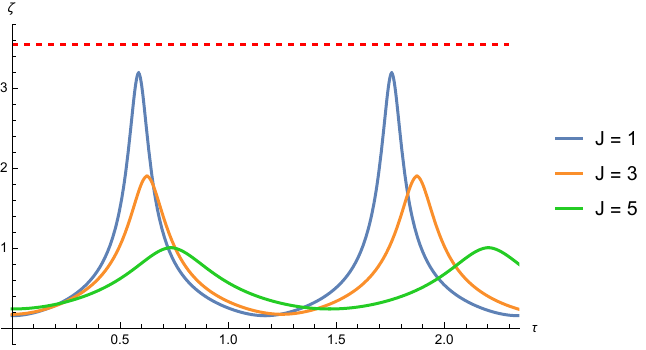}
    	\caption{Particle trajectories for different angular momenta $J$. The end of space is denoted by the red dotted line. The set of parameters used were $(q_1,q_2,q_3,Q^2,\mathcal{H}_{\rm II},J) = (1,2,3,40,7,J)$ and we set $L=1$. The sign of the Hamiltonian depends on the choice of sign in equation \eqref{eqn: zetadot II}. In order to have the end of space $\zeta_*$ real-valued we require $Q^2 L^2 \geq q_1^2 q_2^2 q_3^2$ hence our taking $Q^2=40$.}
        \label{fig: trajectories}
    \end{figure}

    The initial conditions are $\zeta(t = 0) = \zeta_{\text{UV}}$ and $\dot{\zeta}(t = 0) = 0$, the latter of which fixes the value of the energy to be 
    \begin{equation}
    \mathcal{H}_{\rm II} = \frac{m L}{\zeta_{\text{UV}}}\sqrt{G(\zeta_{\text{UV}})^{2}+\frac{J^{2}H(\zeta_{\text{UV}})}{m^{2}f(\zeta_{\text{UV}})}}
    \label{energyII}
    \end{equation}
    for a given choice of $\zeta_{\text{UV}}$ (or vice-versa). Since this equation is difficult to solve analytically for $\zeta_{\rm UV}$ in terms of $\mathcal{H}_{\rm II}$, in practise we choose values for the parameters $(q_1, q_2, q_3, Q^2, \mathcal{H}_{\rm II}, J)$ and then solve $\dot{\zeta}=0$ in \eqref{eqn: zetadot II} numerically to obtain two stationary points. The smaller is the starting position $\zeta_{\text{UV}}$ and the larger corresponds to the turning point $\zeta_0$, in contrast to Section \ref{sec: embedding 1} where the larger always returned the end of space $\zeta_*$. The range over which we solve the differential equation is then $\zeta(t) \in [\zeta_{\text{UV}}, \zeta_0)$. We solve \eqref{eqn: zetadot II} subject to the initial conditions described above. Some example trajectories can be found in Figure \ref{fig: trajectories}. Since they are all possessing non-zero angular momentum $J$ none of the trajectories reach the end of space, and those embeddings with larger angular momentum explore less of the bulk before turning back. We also find that increasing the angular momentum comes with an associated increase in the period of oscillation, which will be inherited by the complexity, as we demonstrate in the following Subsection \ref{sec:Complexity II}.
    
    At early times we can solve \eqref{eqn: zetadot II} order-by-order in $t$, which yields
    \begin{equation}
        \zeta(t) \approx \zeta_{\text{UV}} +\frac{ 2m^{2} f^{2}G \left(G-\zeta_{\text{UV}} G' \right)+J^{2}[f(2H-\zeta_{\text{UV}} H')+\zeta_{\text{UV}} H f']}{4 (m^{2}f G^{2} + H J^{2})  }\bigg|_{\zeta=\zeta_{\text{UV}}}\frac{\zeta_{\text{UV}}t^{2}}{L^{2}} + \mathcal{O}(t^{4}) \, ,
        \label{expzJ0}
    \end{equation}
     where again primes represent differentiation with respect to $\zeta$. Similarly, expanding \eqref{eqn: zetadot II} around the turning point $\zeta_0$ and solving order-by-order yields
    \begin{equation}
        \zeta(t) \approx \zeta_{0} +\frac{ 2m^{2} f^{2}G \left(G-\zeta_{0} G' \right)+J^{2}[f(2H-\zeta_{0} H')+\zeta_{0} H f']}{4 (m^{2}f G^{2} + H J^{2})  }\bigg|_{\zeta=\zeta_{0}}\frac{\zeta_{0}(t-t_0)^{2}}{L^{2}} + \mathcal{O}(t-t_{0})^{4} \, ,
        \label{expztt0}
    \end{equation}
    where we have defined $t_0$ by $\zeta(t_0) = \zeta_0$.
    
\begin{figure}[t!]
    	\centering
    	\includegraphics[width = 6in]{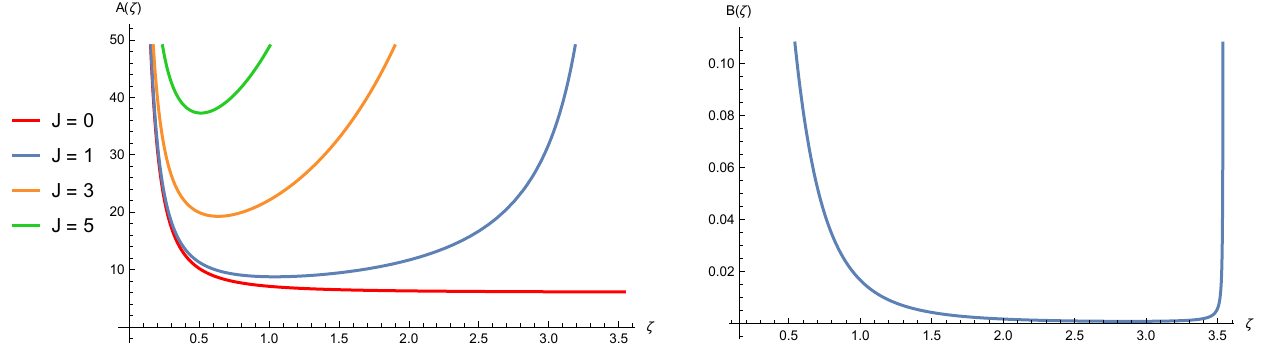}
    	\caption{The effective potential $A(\zeta)$ and effective mass $B(\zeta)$ from \eqref{eqn: AandB}, for the parameter choice $(q_1,q_2,q_3,Q^2,\mathcal{H}_{\rm II},J) = (1,2,3,40,7,J)$. This puts the end of space at $\zeta_* \approx 3.54$. The effective mass has no dependence on $J$.}
        \label{fig: AandB}
    \end{figure}

    As an aside, we mention that the Hamiltonian can be written in the following form:
    \begin{equation}\label{eqn: AandB}
        \mathcal{H}_{\rm II}^2 = A(\zeta) + \frac{P^2_\zeta}{B(\zeta)} \qq \mathrm{where} \qq A(\zeta) = \frac{J^2 H}{f \zeta^2} + \frac{H^{1/3} X_1^{1/2}}{\zeta^2} \qq \mathrm{and} \qq B(\zeta) = \frac{1}{f \zeta^2}
    \end{equation}
    as described in \cite{Nunez:2026vhw}. This means that $A(\zeta)$ can be interpreted as an effective potential for the particle and $B(\zeta)$ as an effective mass, where the $J$ dependence in the former is responsible for the particle's constrained motion in the bulk.  We plot these in Figure \ref{fig: AandB}. At the end of space $\zeta=\zeta_*$ the circle $S^1_\varphi$ shrinks to a point and the angular direction $\varphi$ becomes degenerate. But for any value of $J\neq 0$ the centrifugal barrier in the first term of $A(\zeta)$ diverges, which is responsible for forcing the probe to turn around at the value $\zeta_{0}<\zeta_*$ before it can reach the end of space. Therefore, the geodesic of the particle does not reach the point in the geometry where the circle $S^1_\varphi$ shrinks in this scenario and this procedure happens smoothly. Conversely, when $J=0$ there is nothing preventing the direction $\varphi$ from becoming degenerate and the particle bounces at $\zeta=\zeta_*$ with an impulse which changes its direction instantaneously. This explains the sharp peak of the complexity in the $J = 0$ case, as opposed to the one we calculate below, shown in Figure~\ref{fig: complexity II}.

In the next subsection we study the complexity of this probe.
    
    \subsection{Complexity}\label{sec:Complexity II}  
    \begin{figure}[t!]
    \includegraphics[width= 6in]{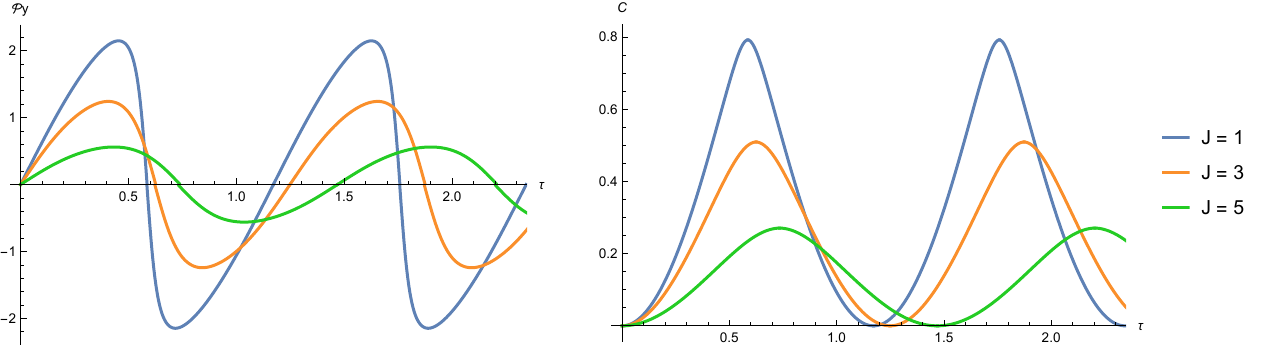}
    \caption{The proper momenta and complexities associated with the particle trajectories in Figure \ref{fig: trajectories}, for the same three $J$ values. We use the Routhian prescription.}
    \label{fig: complexity II}
    \end{figure}
    In analogy with the case in Section \ref{complexityzeta}, we define the proper coordinate for the probe as 
    \begin{equation}
        \dot{y} =  \text{sign}(\dot{\zeta})\, g(\zeta) \frac{L^{2}|\dot{\zeta}|}{\zeta^{2} f(\zeta)^{1/2}} \qq \text{such that} \qq \mathcal{R}=  \sqrt{g(\zeta)^{2}\frac{L^{2}}{\zeta^{2}}-\dot{y}^{2} } \, .
    \end{equation}
    The proper momentum is then given by
    \begin{equation}\label{eqn: proper momentum rou}
        P_{y} =  \frac{|\dot{y}|}{\sqrt{g(\zeta)^{2}\frac{L^{2}}{\zeta^{2}}-\dot{y}^{2} } } =  \text{sign}(\dot{\zeta})\,\frac{ |\dot{\zeta}|}{\sqrt{ \frac{f(\zeta)\zeta^{2}}{L^{2}}-\dot{\zeta}^{2}}} \, .
    \end{equation}
    For this case $[\dot{y}] = \frac{1}{\text{length}}$ which means that $[P_{y}] = 1$. Notice that the proper momentum has the same form as in \eqref{proper0}, however it retains an implicit dependence on $J$, as $\dot{\zeta}$ still depends on $J$.
    Numerical plots for the proper momentum and the complexity are shown in Figure \ref{fig: complexity II}. The first thing to note is that the amplitude of the complexity decreases with increasing $J$, which is in contrast to previous works \cite{Fatemiabhari:2025usn, Fatemiabhari:2026goj}, as a consequence of adopting the Routhian prescription. It should also be noted that while both the proper momentum and complexity are smooth and continuous, decreasing $J$ brings us closer to the results in Figure \ref{complexityzt}. In the $J \rightarrow 0$ limit we recover the discontinuous proper momentum and kinked complexity from the previous section, as explained above.
    \begin{figure}[t!]
    	\centering
    	\includegraphics[width = 4in]{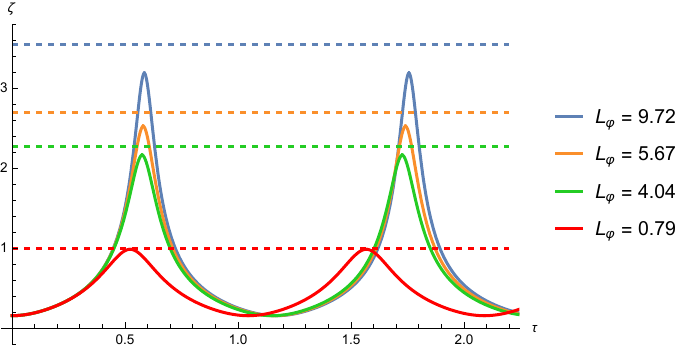}
    	\caption{Particle trajectories for different values of $L_\varphi$, the period of $\varphi$, obtained via varying $Q^2$. The other parameters used were $(q_1,q_2,q_3,\mathcal{H}_{\rm II},J) = (1,2,3,7,1)$. The position of the end of space changes for each parameter choice, and is denoted with a dotted line of the same colour. The trajectories have the same early time behaviour but differ as the particle approaches the end of space.}
        \label{fig: change delta traj}
    \end{figure}
    \begin{figure}[t!]
    \includegraphics[width= 6in]{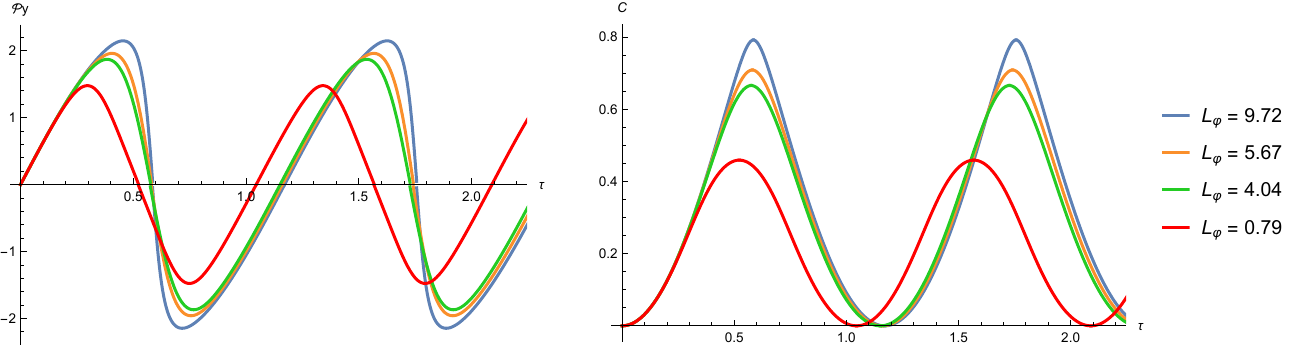}
    \caption{The proper momenta and complexities associated with the particle trajectories in Figure \ref{fig: change delta traj}, for the same four values of $L_\varphi$.}
    \label{fig: Lphi complexities}
    \end{figure}
    
    We can find an early time expansion for the proper momentum by using \eqref{expzJ0}, leading to 
    \begin{equation}
        P_{y} \approx \frac{ 2m^{2} f^{2}G \left(G-\zeta_{\text{UV}} G' \right)+J^{2}[f(2H-\zeta_{\text{UV}} H')+\zeta_{\text{UV}} H f']}{2 L f^{1/2} (m^{2}f G^{2} + H J^{2})  }\bigg|_{\zeta=\zeta_{\text{UV}}} t + \mathcal{O}(t^{3}) \, .
        \label{expansionPyJ}
    \end{equation}
    This is in good agreement with the numerically-obtained solutions. We would like to stress the linear growth behaviour for early times, as it is consistent with the fact that for early times, complexity grows as $\mathcal{C} \sim t^{2}$. We find this behaviour when using the Routhian prescription, which implements the change $\dot{\varphi} \to J$, to a fixed $J$ sector. If we work with the Lagrangian prescription instead, the early time behaviour for the proper momentum is $P_{y} \approx J + c_{1}t^{2}$, not matching the early time behaviour in \eqref{earlytime}. Therefore, in the presence of a Noether conserved charge, one should reduce the system to a fixed charge sector by implementing the Routhian, and from it define the proper momentum.

    By substituting the expansion \eqref{expztt0} in \eqref{eqn: proper momentum rou}, we obtain the behaviour of the proper momentum at the turning point, yielding

     \begin{equation}
        P_{y} \approx \frac{ 2m^{2} f^{2}G \left(G-\zeta_{0} G' \right)+J^{2}[f(2H-\zeta_{0} H')+\zeta_{0} H f']}{2 L f^{1/2} (m^{2}f G^{2} + H J^{2})  }\bigg|_{\zeta=\zeta_{0}} (t-t_{0}) + \mathcal{O}(t-t_{0})^{3} \, ,
    \end{equation}
    where $t_{0}$ is the time at which the probe reaches the turning point. Notice that this expansion diverges when $J = 0$, as in this case we have $\zeta_{0} = \zeta_{*}$, implying $f(\zeta_{0}) = 0$. This is in agreement with Figure \ref{complexityzt}.
    
    In Figure \ref{fig: change delta traj} we plot embeddings with angular momentum $J=1$, but vary the size $L_\varphi$ of the shrinking circle. Recall that $L_\varphi$ is defined from the regularity condition \eqref{eqn: regularity condition}, and as a length scale is related to the confining nature of the dual field theory. We observe that reducing $L_\varphi$ brings the end of space closer to the boundary $\zeta = 0$, and in doing so the particle trajectory reaches a turning point closer to the end of space. Notice also that the period of oscillation decreases with $L_\varphi$.

    In Figure \ref{fig: Lphi complexities} we plot the proper momenta and complexities for the trajectories obtained by varying $L_\varphi$. The amplitude of the complexity decreases with $L_\varphi$, and both proper momentum and complexity inherit the period of oscillation of the corresponding particle trajectory from Figure \ref{fig: change delta traj}, which decreases with $L_\varphi$.

    An interesting feature is that the mechanism by which the particle reverses its radial motion has an impact on the behaviour of the complexity. Compare, for instance, Figure \ref{complexityzt} with Figures \ref{fig: complexity II} and \ref{fig: Lphi complexities}. For $J=0$, the particle reaches the end of space $\zeta_*$, at which point the sign of $\dot{\zeta}$ in \eqref{eqn: zetadot I} reverses -- in practise we impose this by hand. This amounts to an instantaneous change in the particle's velocity, resulting in a discontinuity in the proper momentum and, consequently, a spike in the complexity. For $J\neq0$, by contrast, the centrifugal barrier causes the particle to turn around before reaching the end of space. The reversal of the radial motion is therefore generated dynamically, without the need to impose a change in velocity, resulting in smooth behaviour of the complexity. This demonstrates that the presence of a conserved angular momentum is necessary for smooth behaviour in these finite geometries.
 
\section{Comments on initial states}
\label{sec:field_theory}
In this section we motivate the type of operators that are dual to the probes that we are studying. The geometry at the boundary ($\zeta\to \zeta_{\text{UV}}$) is AdS$_{5} \times S^{5}$, meaning that asymptotically, the dual theory is $\mathcal{N} = 4$ SYM on $\mathbb{R} \times S^{1}\times \mathbb{R}^{2}$ (however the following arguments should equally hold for $\mathbb{R}\times S^1\times S^2$). In \cite{Caputa:2024sux} it is stated that an infalling probe particle is dual to a regulated local operator 
\begin{equation}\label{eq:smeared_operator}
    \mathcal{O}_{\epsilon}^{\Delta}(t, \Omega) = e^{-\epsilon \hat{H}}\mathcal{O}^{\Delta}(t, \Omega)e^{\epsilon \hat{H}} \, ,
\end{equation}
where $\Omega$ specifies the point in $ S^{1}\times S^{2}$, $\Delta$ is the conformal dimension of the operator, $\hat{H}$ is the Hamiltonian operator\footnote{we use a hat to distinguish this from the function $H$ entering the supergravity solution.} and $\epsilon$ is the UV cut-off in the CFT, which for our case is $\epsilon = \zeta_{\text{UV}}$. The validity of this statement holds semi-classically when $\Delta \gg1$ and $\epsilon\ll1$ (of order $1/\sqrt{\Delta}$). Consider an operator at $ t = 0$, denoted by $\mathcal{O}_{\epsilon}^{\Delta}(0, \Omega)$. In order to fix the angular momentum, we need to project onto a fixed angular momentum sector, this is done by  \cite{Berenstein:2019tcs}
\begin{equation}
    \mathcal{O}^{\Delta}_{\epsilon,\ell}(0) := \int \dd \Omega_{3} \, Y_{\ell}(\Omega)\mathcal{O}_{\epsilon}^{\Delta}(0, \Omega) \, ,
\end{equation}
where $\ell$ can be chosen such that the dual particle has angular momentum in one particular direction. This operator insertion is dual to a free falling probe particle with mass $m = \Delta$ and fixed angular momentum $\ell$. For illustrative purposes, let us mention that one such operator in~\eqref{eq:smeared_operator} aligned with the calculation of embedding II could be the BMN-like~\cite{Berenstein:2002jq} BPS operator ${\cal O}^{\Delta}=\text{Tr}\,(Z^J)$ with $\Delta =J\gg 1$, where $Z=\mathcal{X}_1+i\mathcal{X}_2$ with $\mathcal{X}_1,\mathcal{X}_2$ belonging to the six ${\cal N}=4$ SYM scalars. This operator describes a particle moving very fast on an angle of $S^5$ and also on $S^1_{\varphi}$ and is charged under the $\text{U}(1)_R$ symmetry associated with the isometry group of $S^1_{\phi_1}$ and satisfies the semi-classical requirement. However the statements above are not necessarily restricted to this operator or our bulk calculation.

Therefore, we define our initial state to be the one created from this operator acting on the vacuum
\begin{equation}
    \ket{K_{0}} = \frac{ \mathcal{O}^{\Delta}_{\epsilon,\ell}(0) \ket{0}}{\sqrt{\bra{0} \mathcal{O}^{\Delta \dagger}_{\epsilon,\ell}(0)\mathcal{O}^{\Delta}_{\epsilon,\ell}(0)\ket{0}}} \, .
\end{equation}
We propose that this initial state is dual to the probe particle after the Routhian prescription is implemented, so that $J = \ell$. Then, the radial motion of the probe is dual to the unitary evolution
\begin{equation}
   e^{-i\hat{H} t}\ket{K_{0}} = \frac{ \mathcal{O}^{\Delta}_{\epsilon,\ell}(t) \ket{0}}{\sqrt{\bra{0} \mathcal{O}^{\Delta \dagger}_{\epsilon,\ell}(0)\mathcal{O}^{\Delta}_{\epsilon,\ell}(0)\ket{0}}} \, . 
\end{equation}
One can then consider the survival amplitude, to compute the Lanczos coefficients from the moments $\mu_{n} = \frac{\dd^{n}}{\dd t^{n}}S(t) \big |_{t = 0}$. For instance, the first Lanczos coefficients are given by \cite{Balasubramanian:2022tpr}
\begin{align}
     & a_{0} = \bra{K_{0}}\hat{H}\ket{K_{0}} \equiv \langle{\hat{H}}\rangle \, ,\\
    & b_{1}^{2} = \mu_{2}-\langle{\hat{H}}\rangle^{2} \, ,
\end{align}
where $a_{0}$ is the energy of the probe at the initial state \cite{Aguilar-Gutierrez:2025kmw}. On the other hand, $b_{1}$ was computed holographically in the expansion \eqref{expansionPyJ}, so that the following equality should hold
\begin{equation}
    b_{1}^{2} = \Lambda\frac{ 2m^{2} f^{2}G \left(G-\zeta_{\text{UV}} G' \right)+J^{2}[f(2H-\zeta_{\text{UV}} H')+\zeta_{\text{UV}} H f']}{2 L f^{1/2} (m^{2}f G^{2} + H J^{2})  }\bigg|_{\zeta=\zeta_{\text{UV}}}
\end{equation}
after the identification $m = \Delta$, $\ell =  J$ and $\epsilon = \zeta_{\text{UV}}$. Moreover, the second moment can be computed
\begin{equation}
    \mu_{2} = \frac{\bra{0} \mathcal{O}^{\Delta \dagger}_{\epsilon,\ell}(0)(-i \hat{H})^{2}\mathcal{O}^{\Delta}_{\epsilon,\ell}(0) \ket{0}}{\bra{0} \mathcal{O}^{\Delta \dagger}_{\epsilon,\ell}(0)\mathcal{O}^{\Delta}_{\epsilon,\ell}(0)\ket{0}} = b_{1}^{2}+\mathcal{H}_{\mathrm{II}}^{2} \, ,
\end{equation}
    where $\mathcal{H}_{\mathrm{II}}$ is given by equation \eqref{energyII}. One could in principle expand equation $\eqref{expansionPyJ}$ to higher orders, however, by comparing with \eqref{earlytime} it is not possible to extract a recursive relation, since there is a non-trivial dependence of the $a_{n}$ coefficients for spread complexity.

  Let us close this section by mentioning that there is no known exact mapping of the of the Krylov basis $\{|K_n\rangle\} _{n=0}^{n_{\text{max}}}$ in terms of bulk gravitational quantities for cases such as the one studied here. Exact mappings are only known for lower dimensional and more controlled scenarios, like JT gravity where $|K_n\rangle$ is mapped to a fixed chord number~\cite{Rabinovici:2023yex}, or in $\text{AdS}_2$ geometries where it was recently shown in~\cite{Jeong:2026iac} that in the continuum limit, Krylov evolution is mapped to a Klein-Gordon field near the $\text{AdS}_2$ throat. However, given that in our semi-classical approach the initial Krylov state ($n=0$) is dual to the insertion of the massive particle at $\zeta=\zeta_\text{UV}$, and also that the probe's turning point $\zeta_0$ (or $\zeta_*$ in the case of embedding I) is dual to the end of the Krylov chain ($n=n_{\text{max}})$, it would be natural to expect that the Krylov index $n$ is associated to the proper coordinate $y$. Such a relation remains heuristic at this level and it would be interesting to uncover more precise statements about it.

\section{Conclusions}\label{conclusions}
In this work we studied holographic spread complexity of states created by unitary evolving local operators of a deformation of $\mathcal{N} = 4$ SYM. Due to the compactification, the theory effectively flows in the IR to a $(2+1)$-dimensional supersymmetric field theory. In the bulk prescription, these operators are mapped to probe particles that follow geodesics in the supersymmetric branch of the background \eqref{eq:metric_10d}. Such probes experience free falling in the radial direction with $\dot{\zeta}(t = 0) = 0$, starting from a $\zeta_{\text{UV}}$ interpreted as imposing an energy cut-off on the dual theory, towards the value $\zeta_{*}$ which represents the end of space, where the IR of the theory is defined and the geometry closes smoothly.

In \cite{Jiang:2025wpj, Anegawa:2024wov}, it was mentioned that Krylov complexity for simple confining systems exhibits oscillatory behaviour. An important remark pointed out in \cite{Nunez:2026vhw} was that oscillatory behaviour in the complexity does not imply confinement, rather it is a consequence of having a finite geometry. This is further supported by this paper.

Among the different embeddings that one can consider compatible with the equations of motion for the probe, we started by considering a purely radial fall. We numerically solved for the trajectory of the probe which exhibits an oscillatory behaviour bounded by the cut-off $\zeta_{\text{UV}}$ and the end of space $\zeta_{*}$. 
As a consequence of this, since the proper momentum satisfies $P_{y} \sim \dot{\zeta} $, the complexity $\dot{\mathcal{C}} \sim P_{y}$ also shows oscillatory behaviour. 

The second embedding corresponds to a probe particle falling radially and rotating in the confining circle direction $\varphi$. Consistency of the equations of motion require the probe to also have a profile in one of the internal angles of the deformed $S^{5}$. This is a consequence of the twisted geometry, necessary to preserve supersymmetry. This probe is dual to a local operator that is charged under the U(1)$^{3}$ R-symmetry, generated by the three gauge fields. Due to the conserved charge carried by this probe, one needs to implement the Routhian prescription, which at the level of the field theory, is equivalent to restricting the evolution of the state generated by the operator to a fixed charge sector. This is important, as now the motion in $\varphi$ is no longer another independent degree of freedom and instead enters only through an effective mass -- this allows us to treat the probe as falling just radially with a fixed angular momentum. As a consequence of this, the early time behaviour of the complexity matches equation \eqref{earlytime} as required by unitary evolution. 

Another important aspect to highlight for the second embedding is that the conserved charge still plays a role in the radial dynamics, as its presence prevents the probe from reaching the end of the space, but the oscillatory behaviour persists. Interestingly, the presence of the conserved charge makes the bounce of the probe become smooth. Furthermore, it was noticed that for higher values of the conserved charge, the amplitude of the oscillations decrease, in contrast with \cite{Fatemiabhari:2026goj, Fatemiabhari:2025usn} -- this is a consequence of implementing the Routhian to define the proper momentum. In this sense, we argue that the presence of a conserved angular momentum contributes to the well-posedness of the problem in the case of confining solutions with finite geometry and we explicitly considered the case with $J=0$ first to make this more apparent.  

There are several possible future directions that extend the work presented here. While we have focused on geodesics of massive particles, it would be interesting to follow~\cite{Nastase:2026lhz,Chatzis:2026ekd,Chatzis:2026oou} and find embeddings of falling strings and $\text{D}p$-branes in type IIB, dual to extended operators in the field theory, and study the influence of excited internal degrees of freedom. Another intriguing direction would be to consider multiple commuting charges $J_i$ which naturally decompose the Hilbert space into sectors of fixed charges ${\cal H}=\bigoplus _i {\cal H}_{J_i}$, and extend the Routhian prescription to include such cases. This can make contact with results developed in recent literature regarding symmetry-resolved spread complexity~\cite{Caputa:2025mii,Caputa:2025ozd} from the perspective of holography. Finally, considering the rich spectrum of the background used here, it would be insightful to study bulk embeddings dual to operators other than the ${\cal O}^{\Delta}_{\epsilon,\ell}$, which are potentially lighter and less restricted by semi-classical arguments, described in section~\ref{sec:field_theory}, allowing us to study in more depth how the full moduli space of vacua can affect the complexity.

\section*{Acknowledgments} We want to thank various colleagues for their input that improved the contents and presentation of this work. In particular we thank: Carlos Núñez, Nicolò Bragagnolo, Ricardo Stuardo, Dibakar Roychowdhury and Horatiu Nastase. The work of D.C. is supported by the FAPESP grant 2026/02614-6. M.H. has been supported by the STFC consolidated grant ST/Y509644/1. The work of R.T. has been supported by EPSRC Grant EP/Z535175/1 and STFC grant UKRI1787. J.W. is supported by the STFC grant no. ST/Y509644/1.

\bibliography{main.bib}
\bibliographystyle{utphys}

\end{document}